\documentclass[12pt]{iopart}

\usepackage{usebib}
\usepackage{graphicx}
\usepackage{color}
\usepackage{here}

\newcommand{\nminv}{nm$^{-1}$}

\newcommand{\s}{ }
\newcommand{\tm}{T$_{m}$}
\newcommand{\szero}{$S_{0}$}
\newcommand{\sq}{$S(Q)$}
\newcommand{\bq}{$b_{q}$}
\newcommand{\alp}{$\alpha_{P}$}
\newcommand{\vus}{$v_\mathrm{US}$}
\newcommand{\sete}{Se$_{50}$Te$_{50}$}
\newcommand{\rr}{$\rho$}
\newcommand{\cpp}{$C_{P}$}

\begin{document}
\title[SAXS supercool Te]{Experimental Observation of Maximum Density Fluctuation in Supercooled Liquid Te}

\author{Yukio Kajihara$^1$, Masanori Inui$^1$, Kazuhiro Matsuda$^2$ and Koji Ohara$^3$}
\address{$^1$ Graduate School of Advanced Science and Engineering, Hiroshima University, Higashi-Hiroshima, Hiroshima 739-8521, Japan}
\address{$^2$ Department of Physics, Kumamoto University, Kumamoto 860-8555, Japan}
\address{$^3$ Faculty of Materials for Energy, Shimane University, Matsue, Shimane 690-8504, Japan}
\ead{kajihara@hiroshima-u.ac.jp}

\date{\today}

\begin{abstract}
 We performed small-angle X-ray scattering measurements of liquid Te using a synchrotron radiation facility and observed maximum scattering intensity near 620 K in the supercooled region (melting temperature 723 K).
 This result is an experimental observation of the ridge line of the critical density fluctuation associated with the liquid-liquid phase transition that is argued to exist in the supercooled region and verifies the existence of the transition.
 Similar results have been reported for supercooled liquid water, demonstrating the universality of the liquid-liquid transition concept (or inhomogeneous model) in explaining the thermodynamics of these ''anomalous liquids''. 
 This will also contribute to a greater understanding of the physics of liquids, including glass-forming liquids.
\end{abstract}


\maketitle

\section{Introduction}
 Liquid Te has long been known as an ``anomalous liquid'' that exhibits many thermodynamic anomalies:
 its density (\rr) is greatest around its melting temperature (\tm=723 K) \cite{te_density_press,sete_density_tsuchiya};
 the pressure variation of the melting temperature shows a maximum at approximately 1 GPa \cite{meltcurve_max_rapoport};
 the heat capacity (\cpp) reaches a maximum in the supercooled region \cite{te_supercool}; and
 the ultrasonic sound velocity (\vus) shows a minimum and maximum at approximately 620 K and 1070 K, respectively \cite{sete_velocity_tsuchiya}.
 Furthermore, its electrical features are different from those of other liquid metals.
 Although it is a semiconductor in the solid state, it exhibits (poor) metallic electrical conductivity when melted \cite{te_mnm_trans_press}, and its electrical conductivity \textit{increases} as the temperature increases up to approximately 1070 K in the liquid region \cite{te_elect_barrue}.
 The existence of liquid-liquid phase transition (LLT) in the supercooled region and its effect on a wide range of temperature and pressure regimes has long been a major topic of debate regarding these anomalies \cite{sete_inhomo_tsuchiya1985,kajihara_ixs_te}.
 This LLT concept was first introduced to explain the electrical properties as a two-species model \cite{te_elect_johnson} or inhomogeneous transport model \cite{te_elect_cohen}.
 However, at that time, Mott clearly objected to this inhomogeneous transport model, stating that
 ``this regime does not exist in general in disordered materials, though it may in fluids near a critical or convolution point'' \cite{te_inhomo_mott}.
 He did not believe that liquids other than supercritical fluids and phase-separating liquids are inhomogeneous.
 However, the application of this LLT concept is not limited to the electrical properties of liquid Te;
 it can explain the melting curve maxima in the high-pressure region \cite{rapoport1967, meltcurve_max_rapoport}.
 This was later extended by Tsuchiya to understand various properties, including thermodynamics and structure \cite{sete_inhomo_tsuchiya1985}.
 In the case of liquid Te, it has been shown that adding impurities, such as Se, can shift the thermodynamic and electrical properties to the high-temperature side, and these mixtures can be considered as hypothetical supercooled liquid Te \cite{sete_elect_perron, sete_density_thurn, kajihara_saxs_sete}.
 Tsuchiya also applied his model to many Te mixtures \cite{sete_inhomo_tsuchiya1982, gete_tsuchiya2005, tsuchiya_aste} to confirm its validity.
 However, mesoscopic inhomogeneity, which is the essence of the model, has not been experimentally proven for a long time, and results are not conclusive.
 Thirty years later, we succeeded in demonstrating the existence of mesoscopic density inhomogeneity in Te-Se \cite{kajihara_saxs_sete} and Te-Ge mixtures \cite{kajihara_saxs_review} by small-angle X-ray scattering measurements (SAXS) using the SPring-8 synchrotron radiation facility in Japan.
 The temperature dependences of the SAXS intensity and its momentum transfer ($ Q$)-dependent slope, which are important parameters regarding the density inhomogeneity, show maxima near the center of the continuous LLT region.
 The maxima were always near the center of the LLT region, even when the Se concentration and pressure were changed and when the mixing element was changed from Se to Ge.
 That is, it was shown that this inhomogeneity is not caused by mixing or by the type of mixing element but by the LLT of Te itself.
 However, the same SAXS measurements were performed for pure liquid Te, but no significant temperature change in the SAXS intensity associated with LLT was observed above \tm \cite{kajihara_saxs_sete}.
 We concluded that the density inhomogeneity was too small to be detected at temperatures above \tm\s because the real LLT was approximately 100 K below \tm\s \cite{te_supercool}.
 
 Meanwhile, liquid water exhibits thermodynamic anomalies (review \cite{water_review_debenedetti}) similar to those of liquid Te \cite{rapoport1967,similarity_water_te_angell,kajihara_ixs_te,kajihara_llcp_review_jpn,kajihara_ixs_water} and is often referred to as an ``anomalous liquid''.
 The density is greatest at 277 K, which is slightly above \tm (273 K),
 the pressure variation of \tm\s is negative,
 the heat capacity shows a divergent-like increase in the supercooled region, and
 the ultrasonic sound velocity shows a maximum at approximately 350 K.
 Furthermore, models that can be classified under the LLT concept (review \cite{water_2tale}. \cite{water_roentgen,water_llcp_poole, water_llcp_mishima}) have also been proposed independently of the aforementioned model of liquid Te.
 Regarding the LLT concept of liquid water, there has been no clear verification experiment, mainly because the actual LLT is located in the deep supercooled ``no-man's land'' \cite{water_llcp_mishima}, and an actual experiment is nearly impossible.
 However, the latest state-of-the-art X-ray free-electron laser (XFEL) technology has recently made it possible to conduct X-ray measurements in this no-man's land \cite{water_xd_xfel, water_xes_xfel, water_saxs_xfel}.
 In particular, SAXS measurements have shown a maximum of the integrated scattering intensity near 235 K in its temperature variation, and they concluded that this demonstrated the maximum isothermal compressibility or a so-called Widom line, which is strong evidence of the LLT concept \cite{water_saxs_xfel}.
 Cautious comments have been made as to whether or not this really is the maximum \cite{water_saxs_xfel_comment_caupin}, so further verification is necessary.
  
 Therefore, a new high-speed X-ray measuring device was installed at the beamline of SPring-8, where we previously performed SAXS measurements, and we repeated these SAXS measurements of liquid Te in the current study.
 In the case of Te, the actual LLT region is considered to be a metastable supercooled state rather than an unstable no-man's land like that of water.
 In fact, Tsuchiya et al. found the region considered to be the actual LLT using a macroscopic sized (sub-millimeter) sample \cite{te_supercool}.
 The new device has a faster effective measuring speed (1 min) compared to the previous one (approximately 30 min) and is expected to complete SAXS measurements before supercooled liquid Te solidifies. 
  

\section{Results and Discussions}
\subsection{Experimental}
 The experiments were conducted at the beamline BL04B2 of SPring-8 in Japan, as for the previous SAXS measurements \cite{kajihara_saxs_sete, kajihara_saxs_review}.
 The incident X-ray energy was 61.7 keV using Si(2 2 0) crystal reflection as a monochromator, but approximately 7 \% of the high-harmonics of 123.4 keV were mixed.
 In this time, this high-harmonic X-ray was used to investigate the behavior of the larger-$Q$ structure factor, as discussed later. 
 Liquid samples were installed in handmade single-crystalline sapphire cells and heated by Mo heaters.
 The cell was set in a low-pressure vessel, which was filled with He gas at a pressure of approximately 1.5 bar.
 The details of the experimental conditions are presented in the review paper \cite{kajihara_saxs_review}.
 In the present experiment, a two-dimensional detector device using CsI scintillation (Varex XRD 1621AN3) was used \cite{sp8_bl04b2_2018}, although we did use an imaging plate (RIGAKU R-AXIS IV++) in the previous experiments \cite{kajihara_saxs_sete, kajihara_saxs_review}. 
 The new detector has 2048$\times$2048 pixels, with a spatial resolution (pixel size) of 200 $\mu$m. 

\subsection{Results of liquid Te}
\begin{figure}[!h]
	\includegraphics[width=150mm]{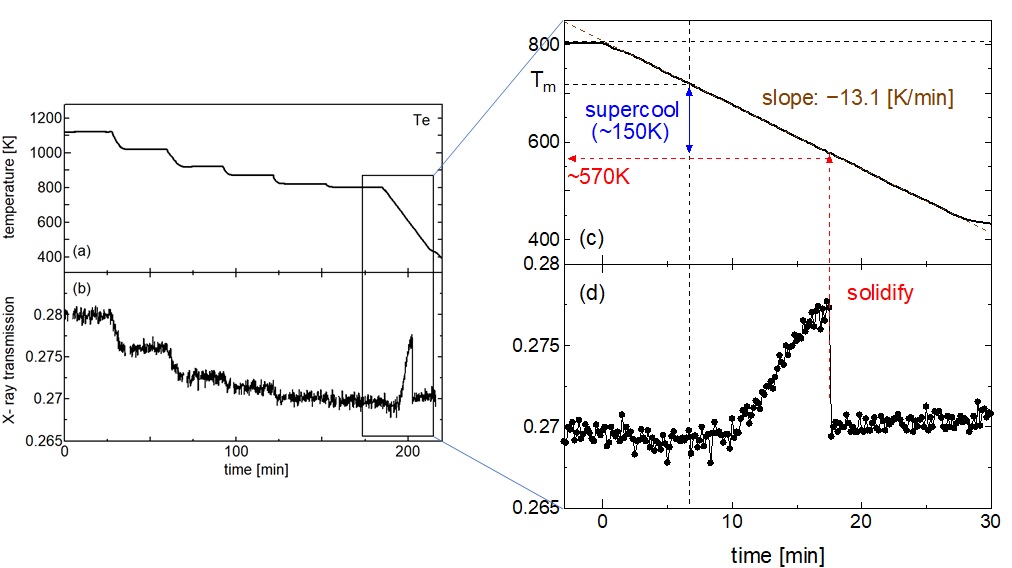}
	\caption{Time evolution of (a) temperature and (b) X-ray transmission of the sample during the measurement. The low-temperature region is enlarged in (c) and (d).}
	\label{fig:time_evolution}
\end{figure}

\begin{figure}
	\includegraphics[width=70mm]{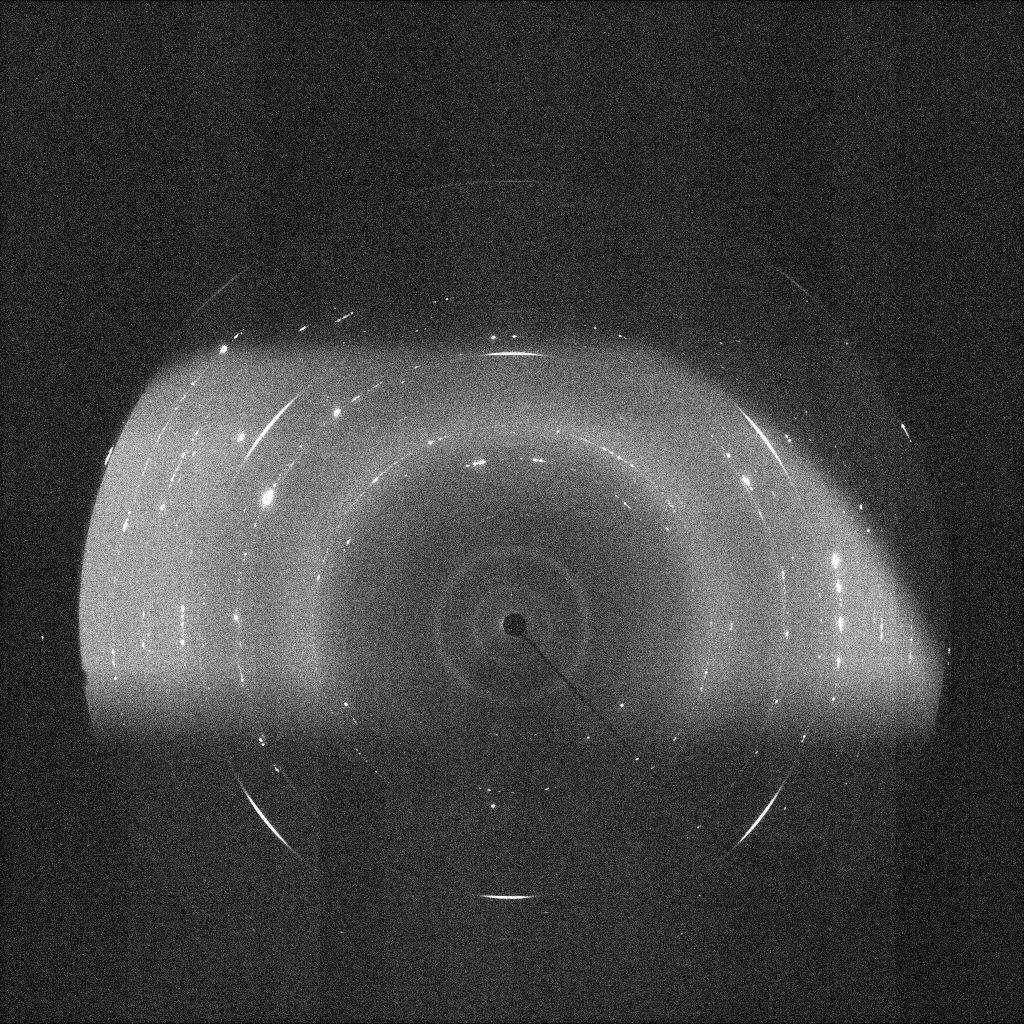}
	\includegraphics[width=70mm]{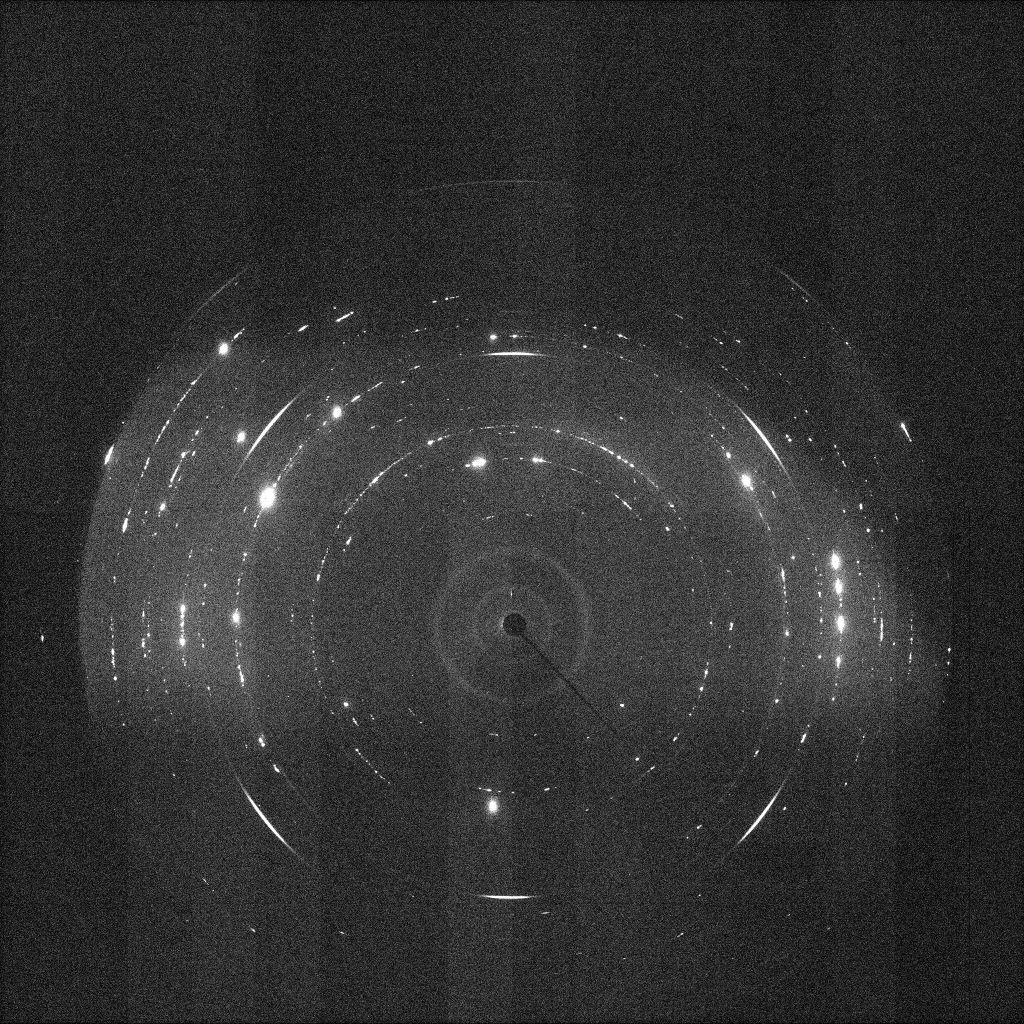}
	\caption{SAXS image of the sample (a) just before and (b) just after the sample solidified.}
	\label{fig:saxs_image}
\end{figure}

\begin{figure}
	\includegraphics[width=150mm]{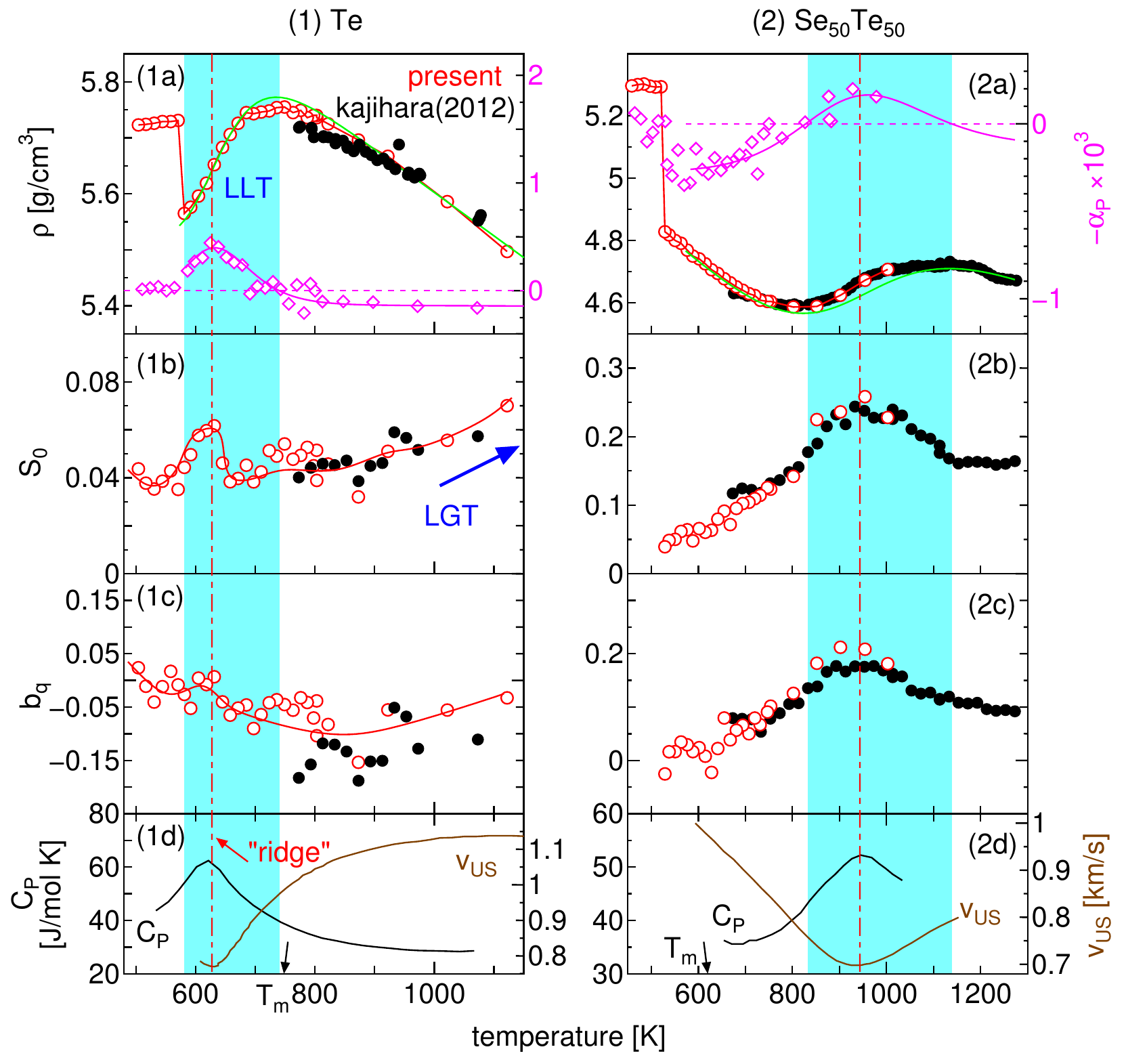}
	\caption{Temperature variation of the obtained (a) \rr\s and SAXS parameters ((b) \szero\s and (c) \bq) from the present and the previous measurements plotted by red open and black closed circles, respectively: (1) for liquid Te and (2) for liquid \sete. The red curves are visual guides. The green curves are those from the literature \cite{te_supercool,sete_density_thurn}. On the right axis of (a), \alp\s calculated from the temperature variation of these densities are plotted by pink marks and lines, respectively. \cpp\s and \vus\s \cite{te_supercool,sete_heatcapacity_kakinuma,sete_velocity_tsuchiya} are plotted in (d) by black and brown curves, respectively. The blue hatched region indicates the LLT region, where \alp\s is negative. The vertical red dot-dashed line indicates the ``ridge line'' of the LLT, where \alp, \cpp\s, and \vus\s exhibit peaks.}
	\label{fig:saxspara}
\end{figure}

 Figure~\ref{fig:time_evolution}(a) shows the time evolution of the temperature during the measurement.
 The X-ray transmission used to determine \rr\s is also plotted in Fig.~\ref{fig:time_evolution}(b).
 After increasing the temperature to approximately 1100 K to melt the sample completely, measurement was performed while lowering the temperature.
 At temperatures much higher than \tm, the detector was exposed for 20 min to capture a scattering image while maintaining constant temperature.
 The temperature and transmission changes were almost always linked (see Fig.~\ref{fig:saxspara}(a)), indicating that the sample responded sufficiently quickly to the temperature change and was also sufficiently stable.
 Meanwhile, in the low-temperature region, including the supercooled region (within the rectangular frame indicated by the dashed lines in Fig.~\ref{fig:time_evolution}(a,b)), the scattering images were captured every 1 min while cooling the temperature at an almost constant rate to prioritize speed.
 Figure~\ref{fig:time_evolution}(c, d) shows an enlarged image of this part.
 As shown in Fig.~\ref{fig:time_evolution}(c), the cooling was performed at an almost constant rate of -13.1 [K/min].
 During this cooling, the transmission (Fig.~\ref{fig:time_evolution}(d)) increased rapidly just below \tm, which corresponds to the fact  that \rr\s reaches its maximum near \tm\s and decreases in the supercooled region \cite{te_supercool} (see also Fig.~\ref{fig:saxspara}(1a)).
 With further cooling, the transmission drops sharply at the red arrow, indicating that the sample has solidified. 
 Below this temperature, the transmission remains almost constant.
 The supercooled state of approximately 150 K was maintained until it solidified. 
 Figure~\ref{fig:saxs_image} shows the SAXS images (a) before and (b) after the sample has solidified.
 (a) Before it solidified, the halo pattern characteristic for liquids was visible;
 however, (b) after it solidified, a spotty peak appeared, indicating that the sample solidified into polycrystalline state.
 Please note that the other patterns visible in both liquid and solid states are due to the sample cell made of single-crystalline sapphire and the Kapton films set in the vacuum path.
 The scattering pattern was basically limited to the rectangular area neat the center due to a slit-shaped optical window located behind the sample.

\begin{figure}
	\includegraphics[width=120mm]{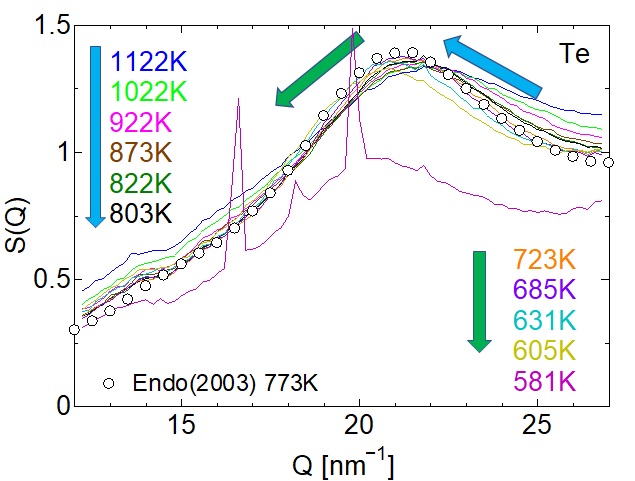}
	\caption{Obtained \sq\s of liquid Te in the large-$Q$ region (solid lines). The temperatures are indicated in the figure. Open circles represent the \sq\s obtained by ND measurement.}
	\label{fig:sq_xd}
\end{figure}

\begin{figure}
	\includegraphics[width=150mm]{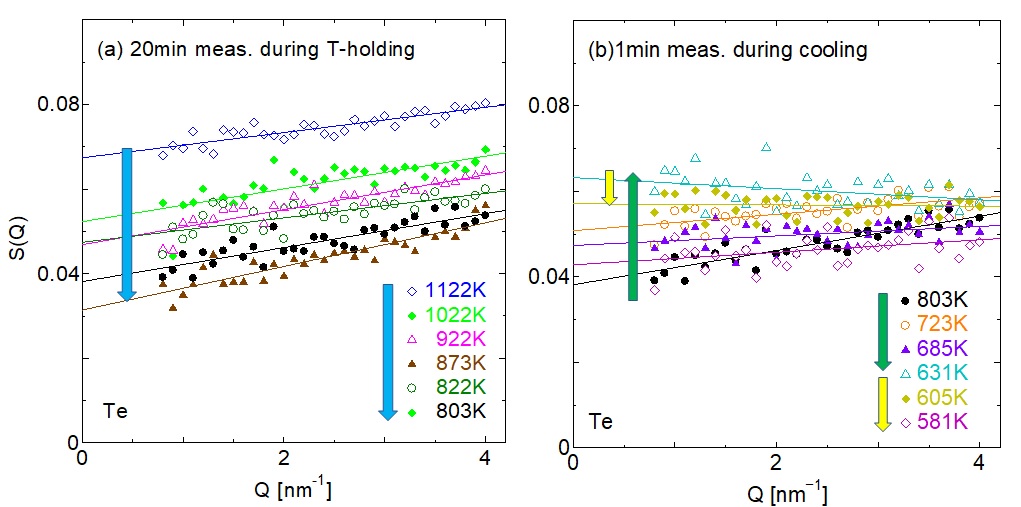}
	\caption{Obtained \sq\s of liquid Te in the small-$Q$ region at (a) high and (b) low temperature. The temperatures are indicated in the figure.}
	\label{fig:sq_saxs}
\end{figure}

 The obtained scattering patterns were carefully analyzed \cite{kajihara_saxs_review} considering scattering from the cell, fluorescent X-ray from the sample, and multiple scattering in the sample, and the $Q$-dependent scattering intensities from the liquid sample were obtained.
 Figures~\ref{fig:sq_xd} and \ref{fig:sq_saxs} demonstrate the temperature variations of these scattering intensities in the large-$Q$ and small-$Q$ regions, respectively. 
 Note that the main peak observed at approximately 20 \nminv\s in Fig.~\ref{fig:sq_xd} is due to the high-harmonic X-ray, which is contained in approximately 7 \% of the incident beam. 
 Because the absolute value of \sq\s was not experimentally determined in the present measurements, the scattering intensities in the large-$Q$ and small-$Q$ regions were separately normalized to fit the \sq\s obtained by neutron diffraction (ND) measurements \cite{te_nd_exafs_endo2003} and the $S(Q\rightarrow0)$ value from our previous SAXS results \cite{kajihara_saxs_sete}, respectively.
 Figure~\ref{fig:sq_xd} shows the result in the large-$Q$ region:
 Over the entire temperature range from 1122 K to solidification at 581 K, the peak position of \sq\s shifted to smaller $Q$, reflecting the anomalous behavior of liquid Te, where the distance to the nearest atomic neighbor increases despite cooling.
 The peak intensity of \sq\s increased gradually up to around the melting point and then decreased slightly.
 The overall shape and temperature changes above \tm\s are close to those of the neutron results \cite{te_nd_exafs_endo2003}, whose \sq\s at 773 K is shown by the open circles in the figure.
 However, in the small-$Q$ region, a different behavior can be observed.
 In the high-temperature region shown in Fig.~\ref{fig:sq_saxs}(a), the intensity decreases almost monotonously, while the slope remains almost unchanged. 
 This temperature variation is normal for liquids.
 However, below this temperature, the slope of the $Q$ dependence changes and $S(Q\rightarrow0)$ increases.
 $S(Q\rightarrow0)$ exhibits the highest value at approximately 620-630 K and decreases again below this temperature.
 To clarify the temperature variation, the scattering intensity was fitted by the linear function $S(Q)=S_{0} (1 -b_{q} Q)$.
 Figure~\ref{fig:saxspara}(1) represents these parameters: (1b) $S_{0}$ and (1c) \bq\s (red open circles).
 We also plot the previous results \cite{kajihara_saxs_sete} above \tm\s as black closed circles.
 Although the slope parameter \bq\s varies significantly, we believe that our results are mostly consistent.
 (1a) $\rho$ estimated from the present X-ray transmission (Fig.~\ref{fig:time_evolution}(b)) is also plotted by the red open circles with a line, which almost agrees with the values in the literature \cite{te_supercool}, indicated by a green line, and the previous results \cite{kajihara_saxs_sete}, indicated by black closed circles. 
 The temperature of 581 K at which the liquid sample solidified is almost the same as in the literature, which indicates that this solidification is due to the spinodal limit or that the sample became unstable with excessive supercooling.
 In Fig.~\ref{fig:saxspara} (1a), the thermal expansion coefficient calculated from the temperature variation of densities \rr\s as $\alpha_{P}=-(1/\rho)(\partial\rho/\partial T)$ from the present experiment and literature values are also plotted by the pink squares and lines on the right axis, respectively.
 In the figure, a negative thermal expansion region between 581 K and 723 K, which is considered to be the LLT region, is indicated by the blue hatch.
 With cooling from 1123 K to approximately 750 K (1b), \szero\s gradually decreased, which can be interpreted as the liquid becoming more homogeneous because it moved away from the critical point of liquid-gas transition (LGT), which should be located at a higher temperature.
 This behavior is more clearly seen for liquid water \cite{kajihara_ixs_water}.
 In contrast, in the LLT region, \szero\s clearly increases to a maximum at approximately 620-630 K.
 This behavior of maximum scattering intensity of supercooled liquid Te was also confirmed in later similar experiment, and more detailed discussions including the large-$Q$ region have been made \cite{te_supercool_saxs_peihao}.
 Around this temperature, (1a) \alp, (1d) isobaric heat capacity $C_{P}$, and ultrasonic sound velocity \vus\s \cite{te_supercool} all exhibit peak values.
 These facts clearly indicate that the maximum of \szero\s is due to the density fluctuation associated with LLT and can be interpreted as the critical ridge line of the LLT or Widom line.
 With further cooling, \szero\s decreases to a smaller value.
 Note that this large change in \szero\s is observed only in the LLT region below \tm, and there is almost no change in the real liquid above \tm.
 It is reasonable that no change was observed in the previous measurement \cite{kajihara_saxs_sete} at temperatures higher than \tm, as indicated by the black closed circles in the figure.
 The parameter \bq\s shown in Fig.~\ref{fig:saxspara} (1c) shows mostly no significant changes above \tm.
 However, in the supercooled region, it seems to show a maximum at approximately 620-630 K, which is in conjunction with the variation of \szero, although the linearly rising tendency is relatively large. 
 Regarding this \bq\s parameter, which corresponds to the domain size of the density inhomogeneity, further investigation may be required.
 In the case of liquid water, it has also been noted that the temperature change in the characteristic size of the density fluctuation is not the usual Ising-like critical behavior \cite{water_saxs_xie}.
 
\subsection{Results of liquid Se-Te mixture}
 We also conducted SAXS measurements for liquid \sete mixture, and the results are shown in Fig.~\ref{fig:saxspara}(2) as red open circles.
 The marks and lines are the same as those for the results of liquid Te shown in Fig.~\ref{fig:saxspara}(1).
 The data in the supercooled liquid region below \tm\s ($\simeq$ 620 K) until solidification at 520 K was newly obtained.
 At temperatures above \tm, the results of the current and previous experiments were close, indicating their reliability.
 (2b) \szero\s shows a maximum near the center of the LLT region indicated by the blue hatch, where (2a) \alp, (2d) $C_{P}$, and \vus\s all show peaks, clearly indicating that the LLT is the cause of the observed density fluctuation.
 By contrast, in the supercooled liquid region ($520 < T < 620$ K), there is no sign of a significant increase in \szero,
which means that cooling or supercooling do not induce the density fluctuation.
 In liquid Te, the temperature at which \szero\s exhibited its maximum was in the supercooled region.
 However, we can conclude that the cause of the density fluctuation is not supercooling but instead the LLT.


\section{Concluding Remaks}
 We performed SAXS measurements for liquid Te and successfully observed that the SAXS intensity shows a maximum, which indicates that the density fluctuation shows a maximum in the supercooled region;
 its temperature variation is linked to that of \rr, \alp, \vus, and \cpp, and it clearly proves that the origin of this density inhomogeneity is not cooling or supercooling but the LLT, which is proposed to be located in the supercooled region.
 This result experimentally verifies the LLT concept in liquid Te \cite{te_elect_johnson,te_elect_cohen,meltcurve_max_rapoport,sete_inhomo_tsuchiya1985,kajihara_ixs_te}, which has been controversial for a long time.
 Meanwhile, it also provides an answer to the simple and essential question by Mott \cite{te_inhomo_mott}:
 liquids can have mesoscopic inhomogeneity except the LGT supercritical region, if they are located in the LLT superciritical region.
 A remaining question would be the actual effect of this LLT located in the supercooled region on the thermodynamics of the real liquid at temperatures higher than $T_{m}$.
 As mentioned in the introduction, past studies have shown that mesoscopic inhomogeneity or two-fluid nature associated with LLT can explain anomalies in liquid Te over a wide temperature and pressure region, including the supercooled region.
 In this sense, the present SAXS results may not necessarily be sufficient to prove the existence of mesoscopic inhomogeneity for real liquid at higher temperatures than $T_{m}$.
 In fact, the situation is almost same for liquid water, the same anomalous liquid, where the two-fluid nature is important for understanding the thermodynamic properties of real liquid above $T_{m}$ \cite{water_roentgen}, but where significant change in density inhomoegneity has actually been measured by SAXS only in the supercooled region  \cite{water_saxs_xfel}.
 In this regard, we believe that among the LLT-derived "fluctuations" that affect actual thermodynamics, the density inhomogeneity that can be measured by SAXS is actually the tip of the iceberg, and that there are hidden elements behind it.
 We have devised and proposed a "dynamical fluctuation'' strength measurement method using sound waves as a method to detect such ``fluctuations" beyond static density fluctuation.
 We have actually applied this method to liquid Te \cite{kajihara_ixs_te} and liquid water \cite{kajihara_ixs_water} and confirmed that such dynamic fluctuations exist over a wide range in the real liquid region above $T_{m}$ for both materials.
 In the case of water, in particular, the measured change in dynamic fluctuation strength is linked to that of the isochoric specific heat over a wide temperature and pressure ranges, strongly suggesting that such dynamical fluctuation is the actual origin of the specific heat anomaly.
 The discussion on the essence of static and dynamical fluctuations is speculative at this point \cite{kajihara_ixs_te,kajihara_ixs_water} and is an issue for future work.
 But in any case, by considering the existence of LLT and the effect of static and dynamical fluctuations associated with it, not just Te and water, we hope that our understanding of the thermodynamics of liquids will be deepened.
 We believe that such a direction is also valid for glass-forming liquids, where "dynamic heterogeneity" is important \cite{hetero_supercool_ediger2000}.


\section*{Acknowledgment}
 This work was supported by JSPS KAKENHI under Grant Numbers 16K05475 and 20K03789. 
  The SAXS measurements were performed at the Synchrotron Radiation Facility of Japan (BL04B2 of Spring-8) with the approval of the Japan Synchrotron Radiation Research Institute (JASRI) (Proposal No. 2017B1426).

\section*{References}

\bibliographystyle{iopart-num}
\bibliography{d:/bibliography_kajihara}

\end{document}